\documentclass[12pt]{article}
\usepackage{amsfonts,amssymb}
\usepackage{graphicx}

\newcommand{\R}[1]{\mathbb{R}^{#1}}

\newcommand{\beq}{\begin{eqnarray}}
\newcommand{\eeq}{\end{eqnarray}}

\def \bi{\bibitem}
\def\){\right)}
\def\({\left( }

\def\be{\begin{equation}}
\def\ee{\end{equation}}
\def\bea{\begin{eqnarray}}
\def\eea{\end{eqnarray}}

\newcommand{\rem}[1]{}

\def\S{{\bf S}}

\def \bi{\bibitem}
\def\np {  {\em Nucl. Phys.} }

\def\ltap{\ \raise.3ex\hbox{$<$\kern-.75em\lower1ex\hbox{$\sim$}}\ }
\def\gtap{\ \raise.3ex\hbox{$>$\kern-.75em\lower1ex\hbox{$\sim$}}\ }

\title{On String Tensions in Supersymmetric 
Gauge Theory
}
\author{
Christopher P. Herzog and
Igor R. Klebanov\\  
Department of Physics, Princeton University\\
Princeton, NJ 08544, USA\\ 
  }
\begin{document}
\setlength{\baselineskip}{16pt}
\begin{titlepage}
\maketitle
\begin{picture}(0,0)(0,0)
\put(325,245){PUPT-2011}
\put(325,260){hep-th/0111078}
\end{picture}
\vspace{-36pt}
\begin{abstract}
In some models of ${\cal N}=1$ supersymmetric $SU(M)$ gauge dynamics
(hep-th/9503163 and hep-th/9707244),
the tension of a string ending on $q$ external
quarks is proportional to $\sin (\pi q/M)$, $q=1,\ldots, M-1$.
In this paper we calculate the ratios of the
$q$-string tensions using the recently derived
type IIB gravity duals of ${\cal N}=1$ SUSY gauge theories. Far in the
IR these gravity duals contain an $\S^3$ with $M$ units of R-R 
3-form flux which, upon S-duality, turns into NS-NS 3-form flux.
The confining $q$-string is described by a D3-brane wrapping an
$\S^2 \subset \S^3$ with $q$ units of world volume flux.
For one of the gravity dual backgrounds (Maldacena-Nu\~nez)
a D3-brane probe calculation
exactly reproduces the $\sin(\pi q/M)$ dependence, 
while for another (Klebanov-Strassler) we find
approximate agreement. We speculate on the connection of the
$q$-string tensions 
with D-brane tensions in the $SU(2)$ WZW model.
\end{abstract}
\thispagestyle{empty}
\setcounter{page}{0}
\end{titlepage}

Consider a gauge theory with gauge group $SU(M)$ and no dynamical
fundamental matter. A Wilson loop in the fundamental representation
corresponds to transporting a probe quark around a closed 
contour $C$.
The classic criterion for confinement is that this Wilson loop
obey the area law 
\be
-\ln \langle W_1(C) \rangle = T_1 A(C)
\ee
in the limit of large area.  
An interesting generalization is to
consider Wilson loops in antisymmetric tensor representations with $q$
indices where $q$ ranges from $1$ to $M-1$. $q=1$ corresponds to
the fundamental representation as denoted above, and
there is a symmetry under $q\rightarrow M-q$ which corresponds
to replacing quarks by anti-quarks. These Wilson loops can be thought of as
confining strings which connect $q$ probe quarks on one end
to $q$ corresponding probe anti-quarks on the other.
For $q=M$ the probe quarks combine into a colorless state (a baryon);
hence the corresponding Wilson loop does not have an area law.

It is interesting to ask how the tension of this class of confining
strings depends on $q$. If it is a convex function,
\be T_{q+q'} < T_q + T_{q'}
\ ,
\ee
then the $q$-string will not decay into strings with
smaller $q$. This is precisely the situation found by 
Douglas and Shenker (DS) \cite{DS} in softly broken ${\cal N}=2$ gauge theory,
and later by Hanany, Strassler
and Zaffaroni (HSZ) \cite{HSZ} 
in the MQCD approach to confining 
${\cal N}=1$ supersymmetric gauge theory: 
\be \label{universal}
T_q = \Lambda^2 \sin {\pi q\over M}
\ ,\qquad q=1,2,\ldots, M-1\ 
\ee
where $\Lambda$ is the overall IR scale.
MQCD uses the 5-branes of
11-dimensional M-theory to engineer theories which 
belong to the same universality class as 
conventional gauge theories \cite{Wit,MQCD}.
Using  these tools it is possible to construct a model with ${\cal N}=1$
supersymmetry and gauge group $SU(M)$ which far in the infrared
exhibits confinement, chiral symmetry breaking and many other 
expected phenomena \cite{MQCD}.
In the UV, however, this model has extra
degrees of freedom compared to the usual 4-d gauge theory; hence there
is no logarithmic running or dimensional transmutation.
The formula 
(\ref{universal}) gives definite quantitative predictions for the ratios of
the string tensions in MQCD, but it is not clear to what extent
these predictions might also be true for 
supersymmetric QCD.
In this paper we rederive this formula using a rather different set-up,
gauge/string duality, suggesting that the formula may be quite
robust.

Quite recently, building on successes of the AdS/CFT correspondence
\cite{jthroat,US,EW},
gravity duals of confining ${\cal N}=1$ supersymmetric gauge theories
have been constructed. In \cite{KS,KT,KN} 
a certain ${\cal N}=1$ supersymmetric
$SU(N+M)\times SU(N)$ gauge theory was studied using $N$
D3-branes and $M$ wrapped D5-branes on the conifold.
If $N$ is a multiple of $M$, then this theory cascades down to
$SU(M)$ in the IR \cite{KS}. The $M$ D5-branes are replaced by $M$ units of
R-R 3-form flux which act to blow up a $\S^3$. As a result the
conifold is replaced by the deformed conifold
\be
\sum_{i=1}^4 z_i^2 = \varepsilon^2
\ .
\ee

The 10-d metric of the KS solution \cite{KS}
is a warped product of $\R{3,1}$ and the
deformed conifold
\be \label{specans}
ds^2_{10} =   h^{-1/2}(\tau)   dx_n dx_n 
 +  h^{1/2}(\tau) ds_6^2 \ ,
\ee
where $ds_6^2$ is the Calabi-Yau metric of the deformed conifold.
Its explicit form in terms of certain angular 1-forms $g_i$ is
\cite{KS,cd,MT,Ohta}
\bea \label{metricd}
ds_6^2 = {1\over 2}\varepsilon^{4/3} K(\tau)
\Bigg[ {1\over 3 K^3(\tau)} (d\tau^2 + (g^5)^2) 
 + 
\cosh^2 \left({\tau\over 2}\right) [(g^3)^2 + (g^4)^2]\nonumber \\
+ \sinh^2 \left({\tau\over 2}\right)  [(g^1)^2 + (g^2)^2] \Bigg]
\ ,
\eea
where
\be
K(\tau)= { (\sinh (2\tau) - 2\tau)^{1/3}\over 2^{1/3} \sinh \tau}
\ .
\ee
At $\tau=0$ the angular metric degenerates into
\be 
d\Omega_3^2= {1\over 2} \varepsilon^{4/3} (2/3)^{1/3}
[ {1\over 2} (g^5)^2 + (g^3)^2 + (g^4)^2 ]
\ ,
\ee
which is the metric of a round $\S^3$ \cite{cd,MT}. 
The additional two directions, corresponding to the $\S^2$ fibered
over the $\S^3$, shrink as
\be {1\over 8} \varepsilon^{4/3} (2/3)^{1/3}
\tau^2 [(g^1)^2 + (g^2)^2]
\ .\ee

{}Far in the infrared (at small $\tau$) the warp factor $h(\tau)$
approaches a constant of order $(g_s M\alpha')^2 \epsilon^{-8/3}$.
Thus, for small $\tau$ the ten-dimensional geometry is 
approximately $\R{3,1}$ times an $\R{3}$ bundle over a round $\S^3$:
\bea \label{apex}
ds_{10}^2  \rightarrow  &{ \varepsilon^{4/3}\over 
2^{1/3} a_0^{1/2} g_s M\alpha'} dx_n dx_n 
+   a_0^{1/2} 6^{-1/3} (g_s M\alpha')
\bigg \{ {1\over 2} d\tau^2  + {1\over 2} (g^5)^2
+ (g^3)^2 + (g^4)^2   \nonumber \\  & + {1\over 4}\tau^2
[(g^1)^2 + (g^2)^2] \bigg \}
\ ,
\eea
where  \cite{CKO}
\be
a_0\equiv
\int_0^\infty d x {x\coth x-1\over \sinh^2 x} (\sinh (2x) - 2x)^{1/3}
\approx 0.71805\ .
\ee
The $\S^3$
is supported by $M$ units of R-R flux $F_3$.
The classical 
supergravity approximation is valid in the limits $g_s\rightarrow 0$,
$M\rightarrow \infty$; $g_s M $ is kept large and fixed.

Another type IIB supergravity solution with similar infrared geometry was
constructed by Maldacena and Nu\~nez (MN) in \cite{MN}. 
The MN background is created by $M$ 
D5-branes wrapped over an $\S^2$. In the UV this solution approaches
a linear dilaton background and the 4-d gauge theory decompactifies
to the 6-d Little String Theory. However, in the IR the
$\R{3,1}\times \R{3}\times \S^3$ geometry with $M$ units of
R-R flux is similar to that found in the KS solution \cite{KS}. 
The similarity suggests that this geometry correctly describes the universality class
of ${\cal N}=1$ supersymmetric $SU(M)$ gauge theory.
The universality of the geometric transition where 
$M$ D5-branes wrapping an $\S^2$ are replaced by an $\S^3$
with $M$ units of R-R flux was elegantly demonstrated by a superpotential
calculation in \cite{Vafa}.
The type IIA mirror of this geometric
transition \cite{Vafa} and its lift to M-theory on a manifold of $G_2$
holonomy \cite{AMV,Ach,AW} have lead to new insights into both gauge theory
and M-theory. In this paper, however, we use the type IIB gravity duals
because there are no instanton corrections here, and also because we
will make use of S-duality.

In the gravity dual the confining $q$-string is described by $q$
coincident fundamental strings placed at $\tau=0$ and oriented along the
$\R{3,1}$.\footnote{Qualitatively similar confining flux-tubes were examined in
\cite{CGST} where the authors use the near horizon geometry of
non-extremal D3-branes to model confinement.} 
In the solution of \cite{KS} both $F_5$ and $B_2$
vanish at $\tau=0$, but it is important that there are $M$ units of
$F_3$ flux through the $\S^3$. In fact, this R-R flux blows up the
$q$ fundamental strings into a D3-brane wrapping an $\S^2$ inside the
$\S^3$.\footnote{A similar effect, that $q$
anti D3-branes blow up in the KS background
into an NS5-brane wrapping an $\S^2$ inside the
$\S^3$, has been considered in \cite{Herman}. Both are particular
examples of the Myers effect \cite{Myers}.}
Although the blow-up can be shown directly, for brevity we 
build on a closely related result of Bachas, Douglas and Schweigert \cite{BDS}.
In the S-dual of our type IIB gravity model, 
at $\tau=0$ we find the $\R{3,1}\times \S^3$ geometry with
$M$ units of NS-NS $H_3$ flux through the $\S^3$ and $q$ 
coincident D1-branes along the $\R{3,1}$. T-dualizing along
the D1-brane direction we find
$q$ D0-branes on an $\S^3$ with
$M$ units of NS-NS flux.  This geometry is very
closely related to the setup of
\cite{BDS} whose authors showed that the
$q$ D0-branes blow up into an $\S^2$. 
We will find the same phenomenon, but our probe brane calculation is
somewhat different from \cite{BDS} because the radius of our $\S^3$ is 
different.

After applying S-duality to the KS solution, 
at $\tau=0$ the metric is
\be {\varepsilon^{4/3}\over 
2^{1/3} a_0^{1/2} g_s^2 M\alpha'} dx_n dx_n + b M\alpha' (
d\psi^2 + \sin^2 \psi d\Omega_2^2 )\ ,
\ee
where $b\approx 0.93266$. 
We are now using the standard round metric on $\S^3$
so that $\psi$ is the azimuthal angle ranging from $0$ to $\pi$.
The NS-NS 2-form field at $\tau=0$ is
\be 
\label{NSfield}
B_2 = M\alpha' \left (\psi -{\sin (2\psi)\over 2} \right)
\sin \theta d\theta\wedge d\phi
\ ,
\ee
while the world volume field is
\be F=-{q\over 2} \sin \theta d\theta\wedge d\phi
\ .
\ee
{}Following \cite{BDS} closely we find that the tension of
a D3-brane which wraps an $\S^2$ located at the azimuthal angle $\psi$ is
\be \sim \left [ b^2 \sin^4 \psi +
\left (\psi -{\sin (2\psi)\over 2} - {\pi q\over M}\right)^2 \right ]^{1/2}
\ .
\ee
Minimizing with respect to $\psi$ we find
\be \label{trans}
\psi -{\pi q\over M} = {1-b^2\over 2} \sin (2\psi)
\ .
\ee
The tension of the wrapped brane is given in terms of the solution
of this equation by
\be  \label{dbi}
T_q \sim b\sin \psi \sqrt{ 1 + (b^2 -1) \cos^2 \psi}
\ .
\ee
Note that under $q\rightarrow M-q$, we find $\psi \rightarrow \pi - \psi$,
so that $T_{M-q}= T_q$. This is a crucial property needed for
the connection with the $q$-strings of the gauge theory.

\begin{figure}
\includegraphics[width=4in]{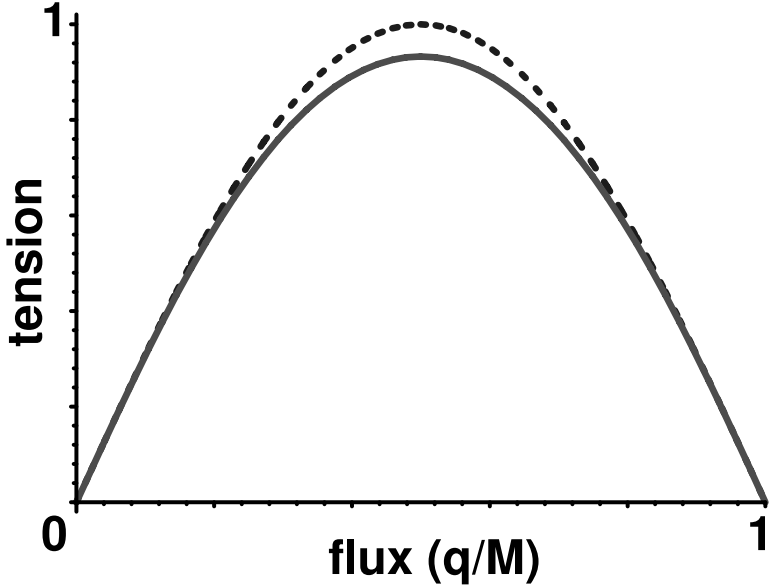}
\caption{A plot of the $q$-string tension versus $q$.  The dashed line
is the MN result ($b=1$).  The solid line is the KS result 
found numerically.
The tension is normalized as in (\ref{dbi}).}
\label{fig1}
\end{figure}

Although (\ref{trans}) is not exactly solvable, we note that
$(1-b^2)/2 \approx  0.06507$ is small numerically. If we ignore
the RHS of this equation, then $\psi\approx \pi q/M$ and
\be \label{probe}
T_q\sim b\sin {\pi q\over M}
\ .
\ee
We have plotted the tension of this $q$-string in Figure 1,
normalized as in (\ref{dbi}).  Note that, even
when $q=M/2$, the tension in the KS case is approximately $93.3\%$ 
of that in the $b=1$ case.

An analogous calculation for the MN background \cite{MN}
proceeds almost identically. In this background only the $F_3$
flux is present; hence after the S-duality we find only $H_3=dB_2$.
The value of $B_2$ at the minimal radius is again given by
(\ref{NSfield}). There is a subtle difference however from 
the calculation for the KS background
in that now the parameter $b$ entering the
radius of the $\S^3$ is equal to 1. This simplifies the probe calculation
and makes it identical to that of \cite{BDS}. In particular, now we
find
\be \label{mainform}
{T_q\over T_{q'}} = {\sin {\pi q\over M}\over \sin {\pi q'\over M} }
\ ,
\ee
without making any approximations.

When we S-dualize back to the original background with RR-flux and $q$
F-strings, all the tensions are multiplied by $g_s$ but their ratios
remain unchanged. Hence, our argument applied to the 
MN background leads very simply to the
DS--HSZ formula for the ratios of $q$-string tensions (\ref{mainform}).
As we have shown earlier, this formula also holds approximately
for the KS background.

It is interesting that the gauge/string duality leads to 
results similar to those of
MQCD, especially as the string tensions are not BPS protected
\cite{Wit}. 
Furthermore, 
we believe that we have strengthened the case
for the robustness of (\ref{mainform}). Since the probe brane is
located at $\tau=0$, our calculation should not be sensitive 
to the UV details of the gravity
dual. The calculation relies only on the IR form of the dual
background where the crucial feature of the geometry is
an $\S^3$ with $M$ units of R-R flux. 

However, we have found that there is
some non-universality in our final result, in that the calculation
with the KS background gives a somewhat different answer than the
calculation for the MN background (only the latter produces
exact agreement with the DS--HSZ formula
(\ref{mainform})).\footnote{Non-universality of this formula in
field theoretic calculations was noted in \cite{Edelstein}.} 
In fact, neither calculation describes the pure ${\cal N}=1$
$SU(M)$ gauge theory.  Our probe brane 
calculations are valid for large $g_s M$, while
to study the pure ${\cal N}=1$ supersymmetric gauge theory we need
to take the limit of small $g_s M$ \cite{KS,MN}.
Hence, we need some control over the $\alpha'$ corrections.
It is interesting that such control can be achieved on the S-dual
side where only the NS-NS flux is present far in the IR (however,
the string coupling becomes strong after the S-duality, so that
the tree level calculation is not reliable).

In fact, the formula (\ref{mainform}) agrees with the exact D-brane 
tensions, including all $\alpha'$ corrections,
in the $SU(2)$ WZW model describing the $\S^3$ with
$M$ units of NS-NS flux \cite{BDS,Aleks,Schweig}. 
In the
level $k$ $SU(2)$ WZW model 
the exact result for the tension of a D3-brane wrapping an
$\S^2$ within $SU(2)$ is \cite{BDS,Aleks,Schweig} 
\be
T_q\sim \sin {\pi q\over k+2}
\label{texact}
\ .
\ee
Making the identification $M=k+2$, which is customary for $M$
NS5-branes, (\ref{texact}) matches exactly the result (\ref{probe}) from
the simple probe analysis in the MN background.

If similar control over the $\alpha'$ corrections could be achieved
in the original sigma model with R-R flux, then
we would learn the exact
ratios of $q$-string tensions in the ${\cal N}=1$ $SU(M)$ gauge theory
from the spectrum of D-branes.
In the meantime we may hope that the 
formula (\ref{mainform}), which as we have shown is
related to exact WZW D-brane tensions, will turn out
to be a good approximation to the gauge theory results
that will one day be extracted from appropriate lattice 
calculations.\footnote{For recent results on $q$-string 
tensions in non-supersymmetric 
gauge theory, see \cite{LT}.}

\section*{Acknowledgements}

We would like to thank T.~Banks,
P.~Ouyang and H.~Verlinde for useful discussions
and A.~Armoni for comments.
This work was supported in part by the NSF grant PHY-9802484.


\end{document}